\documentclass[onecolumn,preprint,preprintnumbers,amsmath,amssymb,floatfix]{revtex4-1}
\usepackage{color}
\usepackage{graphicx}

\begin{document}

\title{Enhanced noise at high bias in atomic-scale Au break junctions}

\author{Ruoyu Chen$^{1}$, Patrick J. Wheeler$^{1}$, M. Di Ventra$^{2}$, D.~Natelson$^{1, 3}$}
\email{natelson@rice.edu}
\thanks{corresponding author}

\affiliation{$^{1}$ Department of Physics and Astronomy, Rice University, 6100 Main St., Houston, TX 77005}
\affiliation{$^{2}$ Department of Physics, University of California, San Diego, California 92093, USA}
\affiliation{$^{3}$ Department of Electrical and Computer Engineering, Rice University, 6100 Main St,.Houston, TX 77005}

\begin{abstract}
Heating in nanoscale systems driven out of equilibrium is of fundamental importance, has ramifications for technological applications, and is a challenge to characterize experimentally.  Prior experiments using nanoscale junctions have largely focused on heating of ionic degrees of freedom, while heating of the electrons has been mostly neglected.  We report measurements in atomic-scale Au break junctions, in which the bias-driven component of the current noise is used as a probe of the electronic distribution.  At low biases ($<$ 150~mV) the noise is consistent with expectations of shot noise at a fixed electronic temperature.  At higher biases, a nonlinear dependence of the noise power is observed.  We consider candidate mechanisms for this increase, including flicker noise (due to ionic motion), heating of the bulk electrodes, nonequilibrium electron-phonon effects, and local heating of the electronic distribution impinging on the ballistic junction.  We find that flicker noise and bulk heating are quantitatively unlikely to explain the observations.  We discuss the implications of these observations for other nanoscale systems, and experimental tests to distinguish vibrational and electron interaction mechanisms for the enhanced noise.
\end{abstract}

\maketitle


Nanoscale junctions are ideal tools for investigating the fundamental processes of heating and dissipation in many-body systems driven out of equilibrium, a situation fraught with complications.  Figure~\ref{Fig.intro} shows some of the relevant lengthscales to consider in this problem.  A ballistic constriction (an atomic-scale junction\cite{Agrait:2003}) bridges between macroscopic source and drain electrodes.   With no bias or temperature gradient applied, the electronic distributions in the source and drain and throughout the junction region are described locally as Fermi-Dirac distributions, characterized by some electrochemical potential (Fermi energy) and a spread in energies given by $k_{\mathrm{B}}T_{0}$, the equilibrium temperature of the system.  Likewise, the local phonon populations (including optical phonons) are characterized by the same temperature, $T_{0}$.  Finally, any local ionic excitations such as the ubiquitous tunneling two-level systems (TLS) undergo dynamics determined by their coupling to the equilibrium phonon and electronic distributions.

When a voltage bias is applied and the system is allowed to reach steady-state, the situation is considerably more complicated, as shown in Figure~\ref{Fig.intro}.  Far from the junction (region A), the widths in energy of the electronic and vibrational distributions remain described by $k_{\mathrm{B}}T_{0}$, presuming that heat is transported away into the bulk effectively.  However, the Fermi energy of the source is raised over that of the drain by an amount $eV$.   In the conventional treatment of such junctions\cite{Datta:1995,DiVentra:2008}, the local electronic distribution functions within regions C and D are no longer simple Fermi-Dirac distributions.  In the drain, for example, right-moving carriers that originated in the source (with energies $\sim eV$ above the Fermi level of the drain) and transmitted through the ballistic region are present.   However, it is generally assumed that those transmitted carriers are drawn from a Fermi-Dirac distribution with a temperature $T_{e} \ge T_{0}$.    We note that dynamical TLS and mobile atoms in region C that couple to electrons are sensitive to the full range of available electronic energies, and hence can respond\cite{Todorov:1998,Montgomery:2002} as if coupled to an effective temperature comparable to $eV$, generally much larger than $T_{0}$ or $T_{e}$.   On a length scale set by the electron-electron inelastic mean free path ($L_{\mathrm{ee}}$), the electrons can exchange energy among themselves, evolving their distribution toward a Fermi-Dirac distribution characterized by a position-dependent, elevated electronic temperature $T_{e} > T_{0}$.  On a still longer distance scale, the electron-phonon inelastic mean free path ($L_{\mathrm{e-ph}}$), the electrons and phonons can both reach approximately thermal distributions given by position-dependent temperatures $T_{\mathrm{e}} = T_{\mathrm{ph}} > T_{0}$.  (The heirarchy of these length scales depends on relative inelastic scattering rates\cite{Giazotto:2006}.  The assumption $L_{\mathrm{ee}} < L_{\mathrm{e-ph}}$ is reasonable in Au at room temperature at a bias of 0.1 V\cite{Groeneveld:1995}, and $L_{\mathrm{ee}}$ is likely to be further reduced in a junction's confined geometry\cite{Roy:2011}.)

In general, it is difficult to define ``temperature'' for a system driven out of equilibrium and indeed several definitions can be advanced~\cite{Dubi:2011}.  In the above, the electronic and vibrational distribution functions in regions A and B are well described as having the thermal functional form, with the effective temperature defined through that description.  In regions C and D, the electronic and ionic populations are not thermally distributed.  However, it is still possible to parametrize such a nonthermal distribution with a single effective temperature.  For example, one can consider allowing local interactions with a large auxiliary ``bath'' (a thermal probe), and determining $T_{\mathrm{eff}}$ as the temperature of the bath such that there is no net steady-state energy flux between the bath and the system~\cite{Dubi:2011}.  Taking an experimental observable (bond strength; current noise; etc.) and comparing that observable with an empirical equilibrium temperature dependence or with a theoretical model containing a temperature parameter is a long-established approach to defining an ``effective'' temperature, though one should be wary of over-interpreting the results.

In the conventional treatment of these systems\cite{Datta:1995,DiVentra:2008}, the electronic distributions \textit{impinging on the ballistic constriction} (region D)  are well approximated as Fermi-Dirac distributions with relative Fermi energies shifted by $eV$.  There can be corrections to this due to back-scattering from disorder in the diffusive source and drain\cite{Ludoph:1999,Ludoph:2000,Untiedt:2000}, and these are modeled as corrections to the overall transmittance.  In the standard picture, the effective electronic temperature $T_{\mathrm{e}}$ of those FD distributions is established by energy redistribution among the electrons via inelastic electron-electron scattering on the scale of $L_{\mathrm{ee}}$ in region B, and by balancing the influx of energy from the bias-driven nonequilibrium distributions and the outflow of energy by thermal conduction of the electrons and eventually the phonons.   The essential work by Henny et al.\cite{Henny:1999} shows that at cryogenic temperatures it is easily possible to get significantly elevated electron temperatures (that is, $T_{\mathrm{e}}-T_{0} > T_{0}$) varying over tens of nanometers.

There have been a growing number of experiments to characterize the energy distributions of the electronic and ionic degrees of freedom in such nanoscale structures.  In the case of ionic degrees of freedom, vibrational occupancies can be inferred directly by highly local Raman spectroscopy measurements\cite{Ioffe:2008,OronCarl:2008,Berciaud:2010,Ward:2011}.  Estimates of effective ionic temperatures may be obtained through measurements of rupture strength of atomic-scale junctions\cite{Smit:2004,Huang:2006,Huang:2007}.  Ionic temperatures may also be determined by measuring the response of an auxiliary system designed to exchange energy with the ions, such as a resistive thermometer\cite{Tsutsui:2012} or a nanoscale thermocouple\cite{Lee:2013}, though these typically indicate effective temperatures tens of nanometers away from the junction.

It is difficult to access the electronic distributions directly, particularly the nonthermal distributions in regions C and D.  These nonthermal electronic distributions have been inferred quantitatively in some elegant experiments in metals\cite{Pothier:1997,Pierre:2000} and carbon nanotubes\cite{Chen:2009,Bronn:2013} via local tunneling measurements.  Noise measurements can provide comparatively direct access to the electronic distribution functions impinging on the constriction (the Fermi-Dirac distributions that go into the nonequilibrium distribution in region D, which dominates the transport).  In equilibrium, the relationship between fluctuations and dissipation leads to the Johnson-Nyquist contribution,\cite{Johnson:1928,Nyquist:1928} current noise that is white over a broad frequency range with a noise power proportional to the absolute (electronic) temperature and the electrical conductance, $S_{I} = 4 k_{\mathrm{B}}T_e G$ A$^{2}$/Hz.  Nyquist-Johnson noise is well established as a primary thermometer.\cite{White:1996}  Additional ``excess''  or ``shot'' noise can appear when a nanostructure is biased out of equilibrium, due to the discrete nature of the electronic charge.  Previous shot noise experiments\cite{Henny:1999} in the diffusive limit, where a metal nanostructure is large compared to the mean free paths for both elastic and electron-electron inelastic scattering ($L_{\mathrm{ee}}$), have found that the Nyquist-Johnson noise and shot noise clearly indicate elevated, bias-driven, position-dependent electronic temperatures $T_{\mathrm{e}}(V) > T_{0}$ in mesoscopic wires at cryogenic temperatures.  Those cryogenic experiments, accompanied by modeling, demonstrate that this heating is a result of the comparatively large thermal resistances of the reservoirs (regions A, B).   We must be mindful of this reservoir heating in any such experiment.

Beyond this quasi-equilibration of the electron subsystem at some effective $T_{\mathrm{e}}$ that takes place in region B, it is an open question whether electron-electron interactions elevate the effective electronic temperature in the immediate vicinity of the junction (regions C and D).  In other words, is the energetic width of the (approximate) FD distributions impinging on the ballistic region D the same as that at the nearby edge of region B, or are there local processes that further broaden those distributions (corresponding to an effective local increase in $T_{\mathrm{e}}$ that is not the reservoir heating described above)?  Theoretical predictions exist regarding local increases of the electronic temperature\cite{DAgosta:2006,Dubi:2011} as a function of bias, but experimental probes have thus far been indirect\cite{Huang:2007,Ward:2011}.  For instance, the bias evolution of a continuum background in anti-Stokes Raman scattering\cite{Ward:2011} ascribed to the electrons has implied effective electronic temperatures elevated by as much as several hundred Kelvin at biases of 0.4~V.  Rupture strengths of molecule-containing junctions, accessing the electrons through their effects on the stability of ionic bonding, were reported\cite{Huang:2007} to be consistent with modeling of electronic heating in a hydrodynamic treatment of the electronic fluid.\cite{DAgosta:2006,Dubi:2011}

For a junction with discrete quantum channels $i$ having transmittances $\tau_{i}$, if the source and drain electronic distributions are thermal with a temperature $T$, the expected form of the excess noise is\cite{Landauer:1991,Martin:1992,Buttiker:1992}:
\begin{equation}
S_{I}=G_0[4k_{\mathrm{B}}T\sum_i^N{\tau_i^{2}}+2eV \coth(\frac{eV}{2k_{\mathrm{B}}T})\sum_{i}^N{\tau_i(1-\tau_i)}]
\label{eq:finiteT}
\end{equation}
where $G_{0} \equiv 2e^2/h$ is the quantum of conductance, including the spin degeneracy.   At zero bias this simplifies to the Johnson-Nyquist noise, while at zero temperature this reduces to:
\begin{equation}
S_{I}=2eVG_{0}\sum_{i}^N{\tau_i(1-\tau_i)}.
\label{eq:zeroT}
\end{equation}
The finite-$T$ expression in Eq.~(\ref{eq:finiteT}) has been confirmed at cryogenic\cite{Cron:2001,Kumar:2012} and room temperature\cite{Chen:2012} in mechanical break junctions.    Often the excess noise is characterized by the Fano factor, $F\equiv \sum_{i}^N{\tau_i(1-\tau_i)}/\sum_{i}^N{\tau_i}$, the ratio of the (zero temperature) bias-driven excess noise to the Schottky case\cite{Schottky:1918} of electrons arriving in Poissonian fashion, $S_{I} = 2eV G$.  

Inelastic processes coupling electrons and local vibrational modes can modify $F$.\cite{Mitra:2004,Chen:2005,Komnik:2009,Avriller:2009,Haupt:2009,Urban:2010,Novotny:2011}  In the limit where electron-phonon scattering can significantly cool the electrons, $F \rightarrow 0$ as in macroscopic conductors.\cite{Blanter:2000}  However, depending on the particular transmittance of channels, $F$ may be either enhanced or suppressed once $eV$ exceeds the level splitting of the local phonon mode, $\hbar \omega$.  These changes have been observed experimentally\cite{Kumar:2012}, though in multichannel junctions unraveling the details can be complicated\cite{Wheeler:2013}.  Depending on whether the local phonon subsystem can equilibrate with bulk phonons quickly, or whether the local phonon population is pumped by the electrons, various dependences of $F$ on the applied bias $V$ are possible.\cite{Haupt:2009,Urban:2010,Novotny:2011}

In this work we report room temperature measurements of the bias dependence of the current noise in scanning tunneling microscope-style Au break junctions, through a lock-in method based on broad band RF detection.  We move beyond previous work\cite{Wheeler:2010,Chen:2012} with a much larger volume of data obtained at biases as much as three times higher, specifically to consider the regime examined in other experiments\cite{Huang:2007,Ward:2011} that report circumstantial evidence of local electronic heating\cite{DAgosta:2006,Dubi:2011} in atomic- and molecular-scale junctions.   At comparatively low biases ($|V| < 150$~mV)  we find bias scaling consistent with excess noise as described in Eq.~(\ref{eq:finiteT}), with an effective electronic temperature equal to the environmental temperature.   At higher biases, the measured noise increases with bias more rapidly than expected from Eq.~(\ref{eq:finiteT}).   A simple, conservative estimate shows that bulk heating of the electrodes, definitely relevant at cryogenic temperatures,\cite{Henny:1999} is comparatively negligible in these 300~K experiments.  We consider flicker noise as a possible contributor to the measured effect and find that this is unlikely to be quantitatively compatible with the observations.  We examine whether the increase in noise could be consistent with nonequilibrium electron-phonon effects and/or estimates of electronic heating local to the ballistic region.  

\section*{Results}

Our experimental approach has been published previously\cite{Wheeler:2010,Chen:2012}  and is described in detail in Methods.  A gold tip is moved repeatedly at a steady rate towards and away from an evaporated gold film, such that atomic-scale Au junctions are made and broken cyclically.  We perform electrical measurements during this process, cycling an applied voltage $V_{0}$ across the junction (and a series resistor) and measuring the corresponding current.  A lock-in approach simultaneously gives the change in integrated radio frequency noise, proportional to $S_{I}(V)-S_{I}(0)$, where $V$ is the voltage drop across the junction.  The radio frequency band employed in the experimental analysis here is from 250 to 600 MHz; data taken with a bandwidth from 400 to 800 MHz is qualitatively identical.   To obtain the ensemble-averaged noise as a function of bias at a given conductance, conductance and noise histograms (500-1000 breaking traces) are acquired at many applied voltage levels.   Figure~\ref{Fig.2} shows a representative conductance histogram, along with ensemble-averaged noise as a function of conductance acquired at several different bias voltages.  The ensemble-averaged data are then binned by conductance and plotted as a function of $V$.   In the picture of Eq.~(\ref{eq:finiteT}), for bias-independent electron temperature and a given junction configuration, plotting  $S_{I}(V)-S_{I}(V=0)$ as a function of the scaled bias $4k_{\mathrm{B}}TG[\frac{eV}{2k_{\mathrm{B}}T}\coth{\frac{eV}{2k_{\mathrm{B}}T}}-1]$, should give a straight line with a slope that is precisely the zero-temperature Fano factor $\sum_{i}^N{\tau_i(1-\tau_i)}/\sum_{i}^N{\tau_i}$.  In our case we expect to find the \textit{ensemble-averaged} Fano factor, $\langle \sum_{i}^{N} \tau_{i}(1-\tau_{i}) / \sum_{i}^{N} \tau_{i} \rangle$.  By fixing the conductance, we have specified $\sum_{i}^{N} \tau_{i}$, so that the slope should be $\langle \sum_{i}^{N} \tau_{i}(1-\tau_{i})\rangle/(G/G_{0})$.

Figure~\ref{Fig.3}(a) shows low bias linear fits at several different conductances away from the peaks in the conductance histograms.  As reported previously,\cite{Chen:2012} small nonlinearities in these scaled plots were observed for conductances close to quantum suppression points, ascribed to extra bias-induced excess noise caused by flicker noise (and possibly some electronic heating of the type discussed below)  while shot noise is minimized at suppressions.

After further increasing bias to higher levels, as shown in Fig.~\ref{Fig.3}(b), significant nonlinearities as a function of scaled bias are present at all the conductance values, well above the extrapolated linear dependence found at low bias.  Here different markers represent different independent data sets while each color indicates a particular conductance value. The solid lines are the low bias linear fits.  In all the data sets, the nonlinearity is concave upward and independent of the order in which the voltages were applied (high to low; low to high; or interleaved).  This indicates that the nonlinearity is an intrinsic function of the bias, not a result of irreversible changes to the junctions during the sequence of histogram acquisition at different applied biases.  The rather similar Fano factors in the several $G_0$ range is consistent with our earlier measurements,\cite{Chen:2012} and may imply the involvement of more quantum channels, suggesting an approach to noise properties similar to those in the diffusive regime.\cite{Beenakker:1992,Nagaev:1992,Liefrink:1994,deJong:1995}


\section*{Discussion}

\begin{figure}[h!]
\includegraphics[width=8cm]{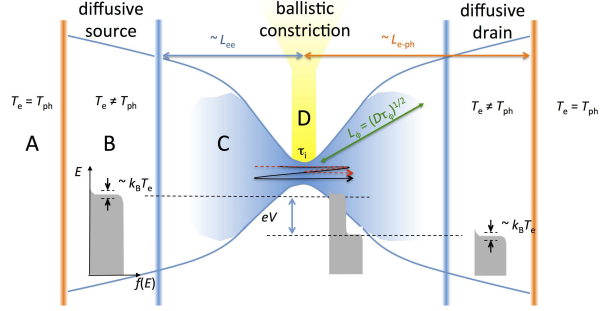}
\caption{A nanoscale junction driven out of equilibrium.  On length scales long compared to the mean free path for inelastic electron-phonon scattering (beyond region A) the electronic and ionic degrees of freedom are assumed to approach approximately thermal distributions parametrized by identical (position-dependent) temperatures (elevated above the equilibrium temperature, $T_{0}$, by an amount depending on the thermal path to ``the environment'').  In region B, the electronic distributions quasi-equilibrate into a local Fermi-Dirac thermal form, parametrized by a position-dependent electronic temperature, $T_{\mathrm{e}}$, greater than that of the ionic degrees of freedom.  In region C, close to the ballistic junction relative to the mean free path for inelastic electron-electron scattering, the electronic distributions are non-thermal, resembling a linear combination of Fermi-Dirac distributions each with some electronic temperature $T_{e}$ and energetically offset from each other by the applied bias, $eV$. Transmission through the ballistic region is through channels of transmittances $\tau_{i}$.  Quantum corrections to the ballistic conductance can arise through quantum interference of trajectories (black and red) involving scattering off disorder in the diffusive electrodes within a coherence length of the ballistic region.  The nonequilibrium electronic distribution in regions C and D can lead to nonequilibrium populations of local phonon modes.}
\label{Fig.intro}
\end{figure}

\begin{figure}[h!]
\includegraphics[width=8cm]{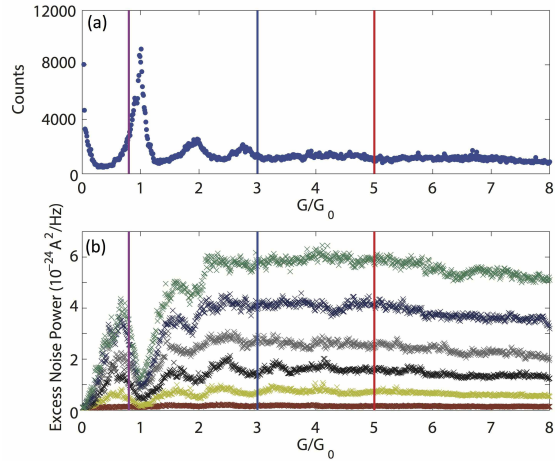}
\caption{(a) An example of a typical conductance histogram as well as the noise measurement. The peaks in histogram indicate conductance quantization, coincident with the relative stability of particular junction geometries. In single valence electron materials like gold, this conductance quantization coincides with suppression of the shot noise, indicating that the chemically stable geometries include significant contributions from fully transmitting channels.  (b) Excess noise as a function of conductance for a series of bias voltages (from bottom to top, 40, 100, 160, 200, 260, and 320 mV).  The vertical lines indicate the particular conductances (purple = 0.8 $G_{0}$, blue = 3 $G_{0}$, red = 5 $G_{0}$) for which noise as a function of bias is plotted in Fig. 3.}
\label{Fig.2}
\end{figure}

At a given conductance, at comparatively low bias, the measured noise scales with bias as expected from Eq.~(\ref{eq:finiteT}).  At comparatively high bias, the measured noise increases more rapidly than this expectation.  We must consider possible explanations for this trend in the noise at high bias.  Possible contributors include: flicker noise; significant heating of the electronic reservoirs (region B in Fig.~\ref{Fig.intro}); electron-phonon inelastic corrections to the noise; and \textit{local} electronic heating (in region C of Fig.~\ref{Fig.intro}).

Flicker noise is produced by fluctuations in the device resistance due to scattering of the electrons by dynamical defects\cite{Hooge:1969,Dutta:1981,Weissman:1988}.  At low bias, such that the dynamical defects remain effectively at $T_{0}$, flicker noise manifests as a voltage noise power that scales quadratically in the applied dc bias.  At higher biases, because of the availablity of carriers as much as $eV$ above the local equilibrium Fermi level, the dynamical defects can have effective temperatures (of their ionic degrees of freedom) that are considerably elevated\cite{Todorov:1998}.  The result of this ionic heating is increased flicker noise as more defects are able to participate, and a bias dependence of the voltage noise that is super-quadratic\cite{Ralls:1989,Holweg:1992}.   The operative question is whether the increased noise we see at high bias is indicative of this kind of enhanced flicker noise contribution.

We assess the role of flicker noise in multiple ways:  considering the overall contribution of flicker noise to the total noise signal; examining the scaling of noise response with RF frequency range; and examining the expected dependence of noise on bias and junction conductance.  The noise measurements in this work are broadband and in the rf regime, a frequency range (250 MHz to 600 MHz) considerably higher than that in which flicker noise is typically measured.  The presence of strong noise suppression at 1 $G_{0}$ at the highest biases (see, e.g., Fig.~\ref{Fig.2}) establishes that the bulk of the measured noise likely results from shot noise of the type described at low bias by Eq.~(\ref{eq:finiteT}).  We have also performed noise measurements with a different RF filter set (limiting the band to between 400 MHz and 800 MHz) at a variety of biases.  While environmental backgrounds proved more annoying over that higher frequency band, the relative magnitude of the noise suppression near values of quantized conductance is essentially unchanged.   This again is consistent with the significant majority of the total measured noise being shot noise, since one would expect overall reduced flicker noise at the higher frequencies.

\begin{figure}[h!]
\includegraphics[width=8cm]{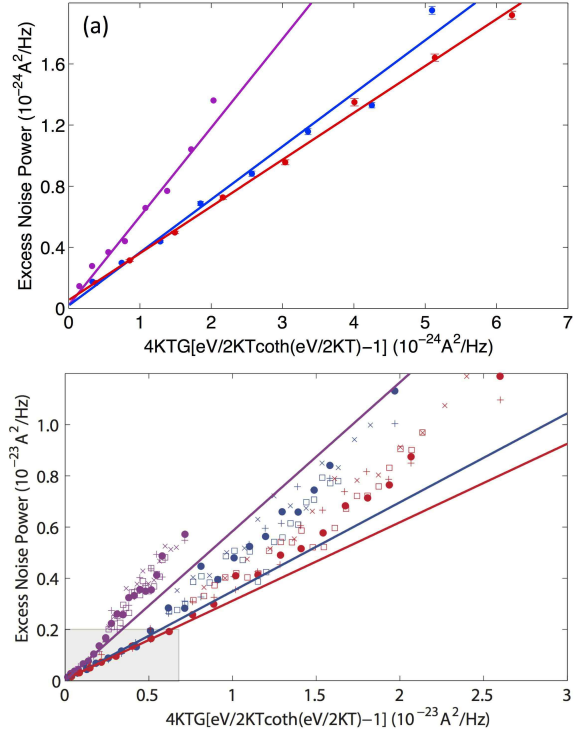}
\caption{Excess noise power vs the scaled bias at (a) comparatively low bias, and (b) comparatively high bias.  The shaded region in (b) is the domain shown in (a).  Colors represent data at different conductances  (purple = 0.8 $G_{0}$, blue = 3 $G_{0}$, red = 5 $G_{0}$), as in Fig. 2. The solid lines are corresponding linear fits.   Ensemble-average Fano factors at 0.8~$G_{0}$, 3 $G_{0}$, and 5 $G_{0}$ are 0.58, 0.35, and 0.32, respectively. (a) At low biases away from the noise suppressions, the excess noise power is quite linear as a function of the scaled bias, consistent with shot noise as in Eq.~(\ref{eq:finiteT}).   (b) At higher biases, significant nonlinearities are present in excess of extrapolations of the low bias linear dependence on the scaled coordinate.  Note that independent data sets (acquired after tip cleaning and annealing) plotted in different markers give consistent results for these nonlinearities.  The filled circle symbols correspond to the data sets in (a) and in Fig. 2(b). }
\label{Fig.3}
\end{figure}

\begin{figure}[h!]
\includegraphics[width=8cm]{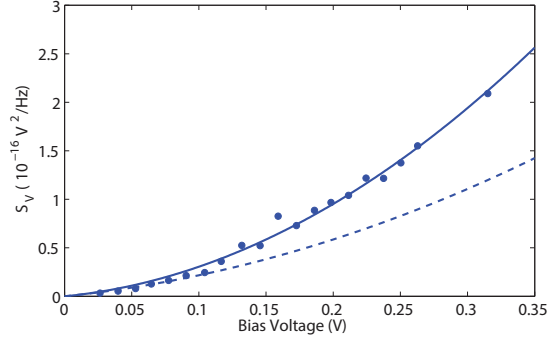}
\caption{Excess noise at 3~$G_0$ converted into voltage noise power $S_V$ as a function of bias.  The solid blue line is the best quadratic fit to the experimental data, while the dotted blue line is a systematic overestimate (see text) of the contribution of flicker noise at 1~$G_{0}$.  Any flicker noise contribution at 3~$G_{0}$ is expected to be even smaller.  Thus, flicker noise is unlikely to be a compatible explanation for the observed increase in noise as a function of $V$.  }\label{Fig.4}
\end{figure}

We look at the bias dependence of the noise.   In previous measurements,\cite{Chen:2012} when looking at 1~$G_{0}$, where the shot noise contribution is maximally (but not completely) suppressed, we found residual nonlinearity with bias in the ensemble-averaged noise power at low bias that was roughly consistent with a flicker noise contribution.  We repeat such an analysis here.  Note that the superlinear behavior in Fig.~\ref{Fig.3}(b) is as a function of \textit{scaled} bias, not just $V$.  Figure~\ref{Fig.4} shows the measured noise (converted to voltage noise, $S_{V}$) as a function of bias across the junction, $V$, for $G= 3 G_{0}$.  The data are well described as a linear + quadratic term.  There is no clear evidence of any superquadratic dependence of the type expected for high bias flicker noise.  

Finally, we consider the expected dependence of noise with the junction conductance.  We can be overly conservative and ascribe \textit{all} of the nonlinearity in $S_{V}$ as a function of $V$ to flicker noise.  This knowingly \textit{overestimates} the magnitude of flicker noise  at 1~$G_{0}$, since the true finite-temperature expression for shot noise (Eq.~(1)) has some nonlinearity due to the $\coth (eV/2k_{\mathrm{B}}T)$ factor.  Previous experimental work at lower frequencies ($< 400$~kHz)\cite{Wu:2008} used lithographically defined mechanical break junctions at cryogenic temperatures to examine flicker noise down to conductances comparable to $G_{0}$ in individual junction configurations.  In those experiments, for a fixed bandwidth, the investigators found that the \textit{ensemble-averaged} flicker noise scales like $S_V/V^2\propto{G^{-1.5}}$ as junctions approach the ballistic regime, though there is variation within the ensemble.   In Fig.~\ref{Fig.4}, we compare the bias-dependent noise at 3 $G_{0}$ with that at 1 $G_{0}$ (dashed line).  Contrary to the ensemble-average scaling expectation for flicker noise, we find that the observed nonlinearity at 3~$G_{0}$ is nearly two times larger than the overestimate of the 1~$G_{0}$ flicker noise.  While large variations are possible between the noise properties of individual junctions,\cite{Wu:2008} this trend (greater nonlinearity vs. $V$  and noise magnitude at higher conductance than at lower conductance) is robust throughout our ensemble-averaged measurements, in both RF frequency bands used, and goes against expectations for flicker noise from previous studies.\cite{Wu:2008}   Thus, based on limits to the overall contribution of flicker noise to the total measured signal, and the systematic noise response as a function of frequency band, bias, and junction conductance, we find it quantitatively unlikely that the increase in noise observed at high bias in Fig.~\ref{Fig.3}(b) is a result of flicker noise.

A second possible explanation for the observed data that must be considered is heating of the ``reservoir'' electrodes due to current flow.   In prior experiments by Henny \textit{et al.}\cite{Henny:1999} considering shot noise at cryogenic temperatures in a diffusive wire connected to thick film reservoirs, the heating of those reservoirs was definitely non-negligible.  In the present experiment, the situation is considerably different:  an atomic-scale junction in what is usually considered the ballistic regime, at \textit{room temperature}, with most of the dissipation happening in effectively bulk reservoirs.  The Au tip has a measured electrical resistance of less than 1~$\Omega$, and the sheet resistance of the Au film is similarly around 2~$\Omega$.   The question is whether power dissipation near the junction could, through the finite thermal resistance of the electrodes, lead to significant heating of the electrodes (both electrons and phonons) near the junction.  To be very conservative, assume that only the electronic component of the thermal conductivity is relevant.   Furthermore, assume that 100\% of the dissipated power is deposited in \textit{either one} of the electrodes proximal to the junction.  For 300~mV bias across a 3~k$\Omega$ junction, the power $P = 3 \times 10^{-5}$~W, and this must correspond to $(T_{res}^{2}-T_{0}^{2}) L G_{res}/2$, where $T_{res}$ is the local temperature of the reservoir electrode near the junction, $T_{0}$ is the environmental temperature, $L = 2.44 \times 10^{-8}$~W~$\Omega$/K$^{2}$ is the Lorenz number, and $G_{res}$ is the conductance of the reservoir, in this case one siemens.  For $T_{0} = 300$~K, we find $T_{res} = 304$~K.  In the real situation, when roughly half of the power will be dissipated in the other electrode\cite{Lee:2013} and phonons will contribute significantly to the thermal path in both electrodes, the actual local temperature rise of the bulk electrodes (region B in Fig.~\ref{Fig.intro}) is expected to be even smaller.  Note that this simple estimate based on the electronic thermal conduction is independent of the detailed electrode geometry.    Therefore, we conclude that this kind of reservoir heating, while definitely important in the cryogenic experiments of Henny \textit{et al.}\cite{Henny:1999}, is unlikely to be a significant contributor to the high bias noise increase in the present work.

At the biases considered in this work, and at room temperature, it is important to consider the possible effects of electron-vibrational coupling on the measured noise.  As mentioned previously, there are many theoretical predictions describing electronic transport in small, ballistic junctions coupled to a local vibrational mode. \cite{Mitra:2004,Chen:2005,Komnik:2009,Avriller:2009,Haupt:2009,Urban:2010,Novotny:2011}  In an individual junction one expects a modification in the Fano factor of the shot noise as the bias exceeds the characteristic energy of the vibrational mode.  The change in $F$ depends sensitively on the particular channel transmittance\cite{Komnik:2009,Avriller:2009,Novotny:2011,Kumar:2012}, as well as the coupling of the channels to the phonon degree of freedom.  

A detailed comparison with particular theoretical descriptions is difficult:  Our data is in the ensemble average, generally with multiple contributing channels; if one expects the Au optical phonon to be a dominant contribution, our data is acquired at the limit $k_{\mathrm{B}}T > \hbar \omega$ so that finite thermal phonon population is possible.   A detailed consideration of ensemble averaging over channel distributions would have to match the observed increase in enhanced noise at higher conductances.  However, the observed power law dependence of the excess noise, $\sim V^{2}$, experimentally constrains possible phonon-based mechanisms for the increase in noise at high bias.  Descriptions that assume strong pumping of the local phonon population by the electrons predict voltage dependences with $V^{3}$ and $V^{4}$ contributions\cite{Urban:2010,Novotny:2011}.  For vibrational effects to be responsible for the enhanced noise, the ensemble-averaged system must be in the ``equilibrated phonon'' limit, such that the local phonon population $\overline{N}$ remains bias independent\cite{Novotny:2011}.  This suggests an experimental test for this mechanism.  By performing such measurements as substrate temperatures are decreased down to the cryogenic regime, one would expect decreases in the local phonon population, and therefore decreases in such nonlinear corrections to the noise. 

Given that temperature gradients within the bulk of the electrodes are small, an additional possible explanation for the observed trends in the measured noise would be some amount of effective electronic heating \textit{local} to the junction (that is, electronic distribution functions impinging locally on the ballistic region (region D) that evolve nontrivially with bias for reasons other than bulk heating of the reservoirs (region B)).  For example, the single-particle scattering formalism neglects electronic heating due to the electronic viscosity, inherently an effect of electron-electron interactions.\cite{DAgosta:2006,Roy:2011}  Before addressing this specific proposal, we attempt to assess whether the observed trend could be consistent with some broadened electronic distribution function (the idea that $T_{\mathrm{e}}$ describing the components of the nonequilibrium distribution function in region C of Fig.~\ref{Fig.intro} is bias-dependent).  
To do this, we parametrize the true nonthermal carrier distribution at the junction location through an \textit{effective} electronic temperature, $T(V)$, that depends on the bias, appearing in Eq.~(\ref{eq:finiteT}).  Obviously, a general nonequilibrium distribution does not have to be parametrized by a single effective temperature; however, some kind of parametrization is necessary to perform an analysis, and this is closely related to the approach of Henny \textit{et al.} when considering ``traditional'' electronic heating.  We employ the Landauer transmittance expression and the resulting Eq.~(\ref{eq:finiteT}), but assume that $T=T(V)=T_0+\delta{T}(V)$, where $T_0 \equiv T(V=0)$ is the ambient temperature, $\approx$~300~K.   To further simplify the analysis, we leverage the fact that $eV\gg2k_{\mathrm{B}}T_{0}$ in the high bias regime of interest.  We can then move forward approximating $\coth{\frac{eV}{2k_{\mathrm{B}}T}}\rightarrow1$, which we will check for self-consistency (to make sure that this continues to hold for the inferred $T(V)$) .   Thus Eq.~(\ref{eq:finiteT}) for a single junction may be reduced to
\begin{equation}
S_{I}(V)=G_0[4k_{\mathrm{B}}(T_0+\delta{T})\sum_i^N{\tau_i^{2}}+2eV\sum_{i}^N{\tau_i(1-\tau_i)}]
\label{eq:finiteV}
\end{equation}
And at zero bias, we already know that
\begin{equation}
S_{I}(V=0)=G_0[4k_{\mathrm{B}}T_0\sum_i^N{\tau_i^{2}}+4k_{\mathrm{B}}T_0\sum_{i}^N{\tau_i(1-\tau_i)}]
\label{eq:zeroV}
\end{equation}
which is purely Johnson-Nyquist noise.  Solving for $\delta T(V)$,
\begin{equation}
\delta{T}(V)=[\frac{S_I(V)-S_I(V=0)}{4k_{\mathrm{B}}G}-(\frac{eV}{2k_{\mathrm{B}}}-T_0)F]\frac{\sum_i^N{\tau_i}}{\sum_i^N{\tau_i^2}}
\label{eq:elT}
\end{equation}
where $S_I(V)-S_I(V=0)$ is exactly the excess noise we measure. The ensemble-averaged Fano factor, $F$, can be determined from the low bias data described above. The only missing information is the ratio $\frac{\sum_i^N{\tau_i}}{\sum_i^N{\tau_i^2}}$.  For our ensemble-averaged measurements, by specifying $G$ we then need to find the ensemble average $\langle \frac{G}{G_{0}\sum_{i}^{N} \tau_{i}^{2}} \rangle$.   While in a single junction with Fano factor $F_{0}$, one can write the parameter of interest as $1/(1-F_{0})$, note that in general one cannot assume that $\langle \frac{G}{G_{0}\sum_{i}^{N} \tau_{i}^{2}} \rangle = 1/(1-F)$, where $F$ is the ensemble average.

\begin{figure}[h!]
\includegraphics[width=8.5cm]{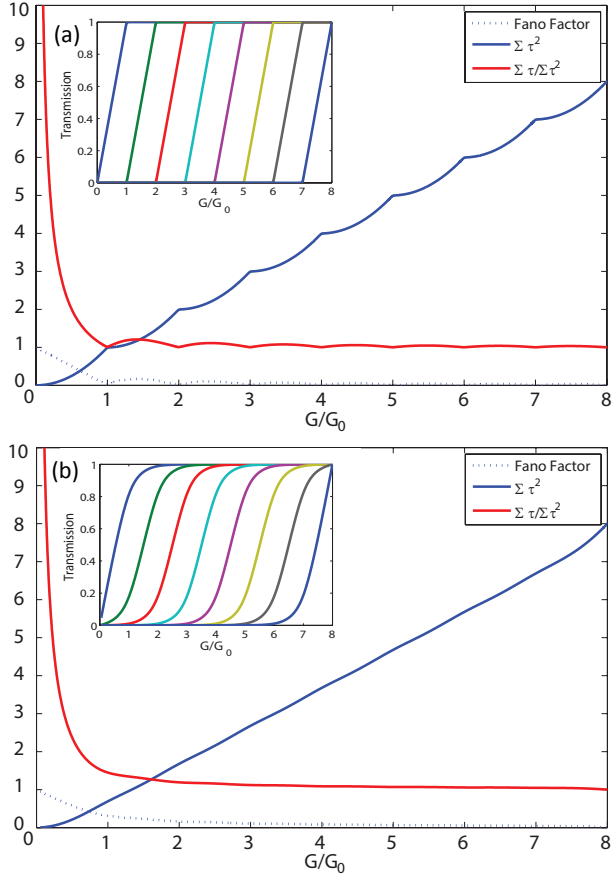}
\caption{A toy model calculation of transport properties.  We consider different models of how particular transport channels with transmittances $\tau_{i}$ contribute to the conductance.   (a) Each channel contributes in turn as conductance is increased, with no channel mixing.  (b) An alternative model, with considerable channel mixing.   In both panels, the dotted lines indicate calculated Fano factors, the solid blue lines denote $\sum_i^N{\tau_i^2}$, and the solid red line shows $\sum_i^N{\tau_i}/\sum_i^N{\tau_i^2}$, all as a function of $G$. }
\label{Fig.5}
\end{figure}

\begin{figure}[h!]
\includegraphics[width=8cm]{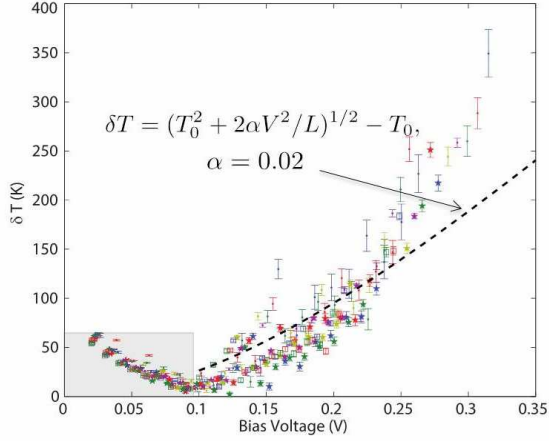}
\caption{Within the model of a locally elevated electronic temperature at the ballistic junction, the inferred electronic temperature from Eq.~(\ref{eq:elT}) vs bias voltage.  For data sets at 13 different conductances {3, 3.25, 3.5, 3.75, 4, 4.27, 4.5, 4.75, 5, 5.25, 5.51, 5.73, 6 $G_0$}, the colors/markers, respectively, are: blue point, red point, green point, purple point, khaki point, blue star, red star, green star, purple star, khaki star, blue open square, red open square, green open square.  (The apparent decrease seen in the shaded region is not physical and results from the failure at low bias of the approximation $\cosh (eV/2k_{\mathrm{B}}T(V)) \approx 1$.)  At high bias, inferred electronic temperatures increase monotonically with $V$, approximately independent of $G$. Error bars are dominated by the statistical uncertainty in the inferred Fano factor of the low-bias data.  Dashed line is a fit to the simple model expression that assumes a local quasithermal electronic distribution, thermal transport by the electrons, and a fraction $\alpha$ of the power $IV$ dissipated locally in the junction, $\delta T = (T_{0}^{2}+ 2\alpha V^{2}/L)^{1/2} - T_{0}$, with $\alpha = 0.02$.  }
\label{Fig.6}
\end{figure}

We can consider models of channel mixing to illuminate possible values, as shown in Fig.~\ref{Fig.5}.  The dotted line represents the Fano factor, the solid blue line $\sum_i^N{\tau^2}$, and the solid red line $\frac{\sum_i^N{\tau_i}}{\sum_i^N{\tau_i^2}}$.  Figure~\ref{Fig.5}(a) represents one limiting case, no channel mixture at all, with each channel turning on linearly sequence as the conductance is increased.  Figure~\ref{Fig.5}(b) assumes much stronger mixing of the channels, so strong that there would be little detectable shot noise suppression in this limit.  In both limits, the ratio of interest approaches a constant (one) quickly after the conductance is increased above 1 $G_0$.  As the diffusive limit is approached,\cite{Henny:1999} $F \rightarrow 1/3$ because of partially open channels, meaning  $\frac{\sum_i^N{\tau_i}}{\sum_i^N{\tau_i^2}} \rightarrow 1.5$.  In either case, it is safe to approximate $\frac{\sum_i^N{\tau_i}}{\sum_i^N{\tau_i^2}} $ as a constant on the order of 1 when $G$ is much above $2 G_0$ (though the precise value introduces a systematic uncertainty in what follows).  Within this approximation, we can evaluate Eq.~(\ref{eq:elT}), and the results are shown in Fig.~\ref{Fig.6}.

The effective electronic temperature elevations in Fig.~\ref{Fig.6} are calculated from the data sets used to produce Fig.~\ref{Fig.3}(b).  In this figure, 13 conductance values are chosen, representative of conductances higher than $G_{0}$.  The approximation $\coth{\frac{eV}{2k_{\mathrm{B}}T}}\rightarrow1$ is only valid outside the shaded region.   At a bias of several hundred mV, within this model an effective electronic temperature elevation as high as 350~K is inferred.   The error bars are dominated by the uncertainty in $F$, as inferred from $\chi^{2}$ of the linear fits in Fig.~\ref{Fig.3}(a).  These local electronic temperature elevations are the same order as those inferred through completely independent experiments employing indirect optical methods\cite{Ward:2011} and bond rupture measurements.\cite{Huang:2007}  

We use a deliberately simplistic model to check the reasonableness of these $\delta T$ values.  The inferred effective temperature elevations are approximately independent of conductance, which is consistent with a local electronic heating picture assuming a balance between local power generation and heat current carried by the electrons.\cite{DAgosta:2006}  In this local heating picture, some fraction $\alpha$ of the total Joule heating is dissipated local to the junction, $P_{\mathrm{in}} = \alpha IV = \alpha V^{2}G$.  (In the standard approach, $\alpha$ would be zero; given the general success of the standard approach, the expectation is that $\alpha$ should be small compared to one.)  Since the electron-phonon inelastic mean free path is much longer than the junction size, the electrons must carry away the heat generated.   In the limit of local quasithermal energy distributions (so that a local electronic temperature is well defined) for the outgoing carriers, the electronic thermal conductance is $L T G$.  Balancing the Joule heating with the thermal current $P_{\mathrm{out}} = (T^{2}-T_{0}^{2})L G/2$ leads to a predicted local temperature of $(T_{0}^{2}+ 2\alpha V^{2}/L)^{1/2}$.  The observed data are roughly consistent with this for $\alpha \approx 0.02$, not bad considering the simplicity of the model, which assumes the Wiedemann-Franz law and no dependence of the $\tau_{i}$ on bias.  The inferred $\delta T$ seems to increase more rapidly than this dependence at higher bias; this simple model may be inadequate, assuming for example that the transmittances $\tau_{i}$ are independent of bias (generally not a poor assumption for Au over this energy range\cite{Ward:2010,Lee:2013}).

We see that the observed noise trend could be explained by a relatively modest amount of electronic heating local to the junction (dissipating 2\% of the total power in region C of Fig.~\ref{Fig.intro}, though in general one would expect $\alpha$ to depend on the conductance).  We now consider whether there is a physical mechanism that could contribute this amount of local electronic heating.  Indeed, similar to the local, nonequilibrium ionic heating due to electron-phonon interactions, the internal Coulomb friction of the electron liquid is predicted to create local electron heating \cite{DAgosta:2006}.  Due to the large current densities in nanojunctions compared to their bulk counterparts the electron-electron scattering rate increases locally in the junction~\cite{DiVentra:2008}. The
underlying Fermi sea thus ``heats up'' locally due to this increased scattering,
via production of electron-hole pairs whose energy needs to be dissipated away from the junction.  This would manifest itself in the noise just as an elevated electronic temperature does so in the regime of Henny \textit{et al.}\cite{Henny:1999}.  The local effect was first computed in Au nanojunctions\cite{DAgosta:2006} assuming no dependence of the electronic viscosity on the junction geometry, namely on the electron transmission. Using the viscosity of that work, namely 
$10^{-7}$ Pa-sec for the electron liquid of density comparable to gold ($r_s=3$), we obtain an estimate of $\delta T$ that is about two orders of magnitude smaller than what is shown in Fig.~\ref{Fig.6}.  The drawback of neglecting the dependence of viscosity on junction geometry was however understood and corrected in Roy \textit{et al.}\cite{Roy:2011}, where the zero frequency viscosity $\eta(0)$ was evaluated as the product of the high-frequency shear modulus $\mu_\infty$ and a junction-specific momentum relaxation time $\tau$. The actual scattering potential used in Ref.~\cite{Roy:2011} to compute $\tau$ was chosen precisely to model Au quantum point contacts.  By referring to Fig.~\ref{Fig.4} of Ref.~\cite{Roy:2011} we see that $\eta(0)\sim 10^{-4}$ Pa sec for the transmissions typical of these point contacts.  Since the zero-frequency viscosity enters as the square root of its value in the effective electronic temperature (see Ref.~\cite{DAgosta:2006}) we obtain a value of $\alpha=0.05$, a factor $2.5$ higher than the one fitted in Fig.~\ref{Fig.6}, a very reasonable agreement considering that the actual geometry of the junction is not known.  This quantitative consistency between the modeled experiment and this theoretical treatment suggests that the electronic viscosity should definitely be considered as a potential contributor to the observed enhanced noise.   Within this theoretical picture the effective electronic heating depends strongly on the transmittance, implying that such heating would be much less important in lower conductance structures such as molecular junctions, where it might be detected through inelastic electron tunneling spectroscopy linewidths.\cite{Hipps:1989}.  We further note that this electron-electron interaction effect should be comparatively independent of the background equilibrium temperature.  Therefore, measurements of the scaling of noise at high bias at lower substrate temperatures should allow the discrimination between electron-vibrational enhancement of the noise discussed above and this local electronic heating mechanism.

In conclusion, we have conducted excess noise measurements in scanning tunneling microscope-style Au break junctions at room temperature and observed a nonlinear bias dependence of the noise exceeding the expectations for shot noise in the Landauer formalism at fixed electron temperature.  The magnitude of this extra noise is inconsistent with flicker noise or heating of the bulk electrodes as possible confounding effects.  Nonequilibrium electron-vibrational effects could be relevant, but the observed power law of the noise increase suggests that any participating local phonon modes are not pumped significantly by the electrons.  Interpreting the increased noise in terms of a model of local electronic heating, we find that the additional noise implies an effective electron temperature (defined through the width of the FD distributions impinging on the ballistic junction) elevated by as much as 300~K at 0.3 V bias.   A theoretical treatment accounting for the effects of electronic viscosity is shown to be roughly consistent with such a value.  It is important to note again that this noise method is only one way to parametrize the local nonequilibrium electronic distribution by an effective temperature.   The temperature determined this way may not be identical to a local, nonequilibrium temperature defined\cite{Dubi:2011} through the lack of net average transport of energy between a reservoir and the nonequilibrium system.   Additional experiments looking at electronic heating as a function of environmental temperature and transmittance should further constrain theoretical models of this nonequilibrium phenomenon, and make it possible to distinguish between electron-vibrational and electron-electron contributions to the enhanced noise.

\section*{Methods}

\subsection*{Measurement technique}

\begin{figure}[ht!]
\includegraphics[width=8cm]{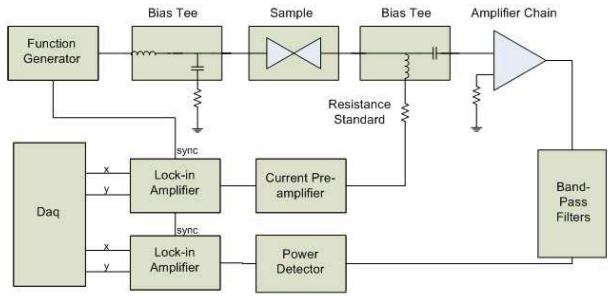}
\caption{Schematic of rf noise measurement scheme.}
\label{Fig.7}
\end{figure}

A schematic of the measurement system is shown in Fig.~\ref{Fig.7}.  A function generator at 10 kHz cycles the applied bias between zero and a particular level $V_{0}$.  This low frequency bias passes through a bias-tee and is applied to the tip; the film is connected, via the DC port of another bias-tee, through a resistance standard (as a current limiter) set to $R_{0}=$~2~k$\Omega$, and then to a current preamplifier.    The current preamplifier output is measured directly by a high speed data acquisition (DAQ) card as well as feeding into a lock-in amplifier synced to the 10 kHz bias.  Knowing $R_{0}$ and the current, we calculate the true voltage drop across the junction, $V$. The high-frequency port of one bias-tee is terminated at 50~Ohms, while the other bias-tee's high frequency port is connected to an amplifier chain, a filter set to define the bandwidth (either 250 MHz to 600 MHz, or 400 MHz to 800 MHz), and then to a logarithmic power meter.  The power meter output is read directly by the DAQ and is also directed into a second (``noise'') lock-in amplifier synced to the low frequency bias.   This second lock-in measures the component of the radio frequency noise (integrated over the gain-bandwidth product of the system including the amplifier chain) due to the applied ``DC'' bias.   The noise lock-in signal (measuring the mean square voltage signal on the power detector) includes a background (independent of junction conductance and bias conditions) due to amplifier noise that may be subtracted away from the ensemble-averaged power measurements.   Measurements of the junction's RF reflectance over the bandwidth as a function of conductance are essentially featureless between 0.1~$G_{0}$ and 8~$G_{0}$, showing that the efficiency of the out-coupling of RF noise from the junction varies little over that conductance range.



\section*{Acknowledgments}
D.N. and R.C. acknowledge support from NSF DMR-0855607 and DMR-1305879.  M.D. acknowledges partial support from DOE under Grant
No. DE-FG02-05ER46204.

\section*{Author Contributions}
RC performed all measurements and analyzed the data. PJW helped develop the measurement setup and the analysis framework and software.  DN supervised and provided continuous guidance for the experiments and the analysis.  MD provided valuable theoretical guidance and insight.  All authors discussed the results and contributed to manuscript revision.

\section*{Competing Interests}
The authors declare that they have no competing financial interests.

\section*{Correspondence}

Correspondence and requests for materials should be addressed to D.N.

\clearpage

\section*{Figures} 

\clearpage

\clearpage

\clearpage

\clearpage

\clearpage

\clearpage

\end{document}